# Characterising Water Exchange in Gliomas Using Diffusion MRI with Free Gradient Waveforms

Arthur Chakwizira[1,3], Filip Szczepankiewicz[2], Carl-Fredrik Westin[3], Linda Knutsson[4,5], Pia C Sundgren[1,6], Markus Nilsson[1]

1. Department of Diagnostic Radiology, Clinical Sciences Lund, Lund University, Lund, Sweden
2. Department of Medical Radiation Physics, Lund University, Lund, Sweden
3. Department of Radiology, Brigham and Women's Hospital, Harvard Medical School, Boston, MA, United States
4. Department of Neurology, Johns Hopkins University School of Medicine, Baltimore, MD, United States
5. F. M. Kirby Research Center for Functional Brain Imaging, Kennedy Krieger Institute, Baltimore, MD, United States
6. Centre for Medical Imaging and Function, Skane University Hospital, Lund, Sweden

## Corresponding author:

Arthur Chakwizira
Department of Radiology, Brigham and Women's Hospital, Harvard Medical School, Boston, MA, United States
Email address: achakwizira@bwh.harvard.edu

## Sponsors/Grant numbers:

VR (Swedish Research Council)

- 2024-04968
- 2023-02412

eSSENCE

- 10:5

Multipark

- This study was supported by MultiPark - A Strategic Research Area at Lund University

ALF

- The study was financed by Swedish governmental funding of clinical research (ALF).

Cancerfonden (The Swedish Cancer Society)

- 22 2011 Pj
- 22 0592 JIA
- 2022/2414
- 2024 24 3568 Pj 01 H
- 2025/4688

Hjärnfonden (The Swedish Brain Foundation)

- FO2024-0335-HK-73

National Institutes of Health

- R01NS125781

**Keywords:** Diffusion MRI, time-dependence, restricted diffusion, exchange, glioma, free waveforms

**Abstract**

Transmembrane water permeability, which regulates cellular water exchange and is influenced by water channels such as aquaporin-4 (AQP4), has been implicated in glioma progression and may affect tumour infiltration and treatment response. Non-invasive mapping of water exchange may therefore provide biomarkers of glioma pathology. This study investigates the feasibility of characterizing water exchange in gliomas using diffusion MRI with free gradient waveforms, known as the Restriction-Exchange (ResEx) approach, which enables exchange quantification independent of restricted diffusion effects. Thirteen patients with histologically confirmed gliomas (ten glioblastomas, three astrocytomas) underwent preoperative MRI at 3T using a custom ResEx protocol. Multiple diffusion-weighted acquisitions with selective exchange sensitivity were performed to estimate voxel-wise maps of the apparent diffusion coefficient (ADC), diffusion kurtosis, and water exchange rate. ResEx-derived maps revealed heterogeneous spatial patterns across and within tumours. Elevated exchange rates were commonly observed in enhancing tumour margins, potentially reflecting smaller cells, increased membrane permeability or AQP4 upregulation. In some cases, elevated exchange extended into non-enhancing peritumoural regions. Exchange values in oedema were slightly higher than in healthy tissue, suggesting potential infiltration or membrane disruption. Diffusion MRI with free gradient waveforms permits non-invasive mapping of water exchange in gliomas and reveals physiological information not captured by standard imaging. Exchange rate mapping may offer novel biomarkers of tumour aggressiveness, infiltration, and treatment response, and holds promise for surgical and radiotherapy planning.

## 1 Introduction

Gliomas are the most common tumours of the central nervous system in adults, accounting for more than two thirds of all primary brain tumours (Omuro & DeAngelis, 2013). Regardless of the intervention used to manage them, gliomas are associated with a poor prognosis—a five-year survival rate of only 5% for the high-grade variants (Nayak & Reardon, 2017; Ostrom et al., 2014). There is mounting evidence that a critical factor determining the fate of gliomas is the membrane channel aquaporin 4 (AQP4), responsible for selectively transporting water molecules across the cell membrane (Jia et al., 2023; Nico et al., 2009; Papadopoulos & Verkman, 2013; Sun et al., 2020). AQP4 is the main water channel of the central nervous system and is normally distributed in the perivascular astrocytic end-feet. In gliomas, AQP4 is upregulated and undergoes redistribution from the astrocytic end-feet to the entire cell membrane (Montgomery et al., 2020; Papadopoulos & Verkman, 2013). It has been shown to play a central role in tumour migration, proliferation, and treatment response. AQP4 expression could thus serve as a prognostic biomarker for gliomas.

A biomarker of AQP4 expression derived from magnetic resonance imaging (MRI) is an appealing prospect, however, AQP4 is challenging to directly image due to its low molar concentration in the brain, making it invisible to both MR spectroscopy and chemical exchange transfer (Blocher et al., 2011; Castañeyra-Ruiz et al., 2013; Jia et al., 2023). A potential alternative is to measure the transmembrane water exchange rate, which is related to AQP4 expression (Verkman et al., 2008; Yang & Verkman, 1997). An excellent approach for mapping membrane permeability is diffusion MRI, due to its inherent sensitivity to tissue microstructure (Andrasko, 1976; Kärger, 1985). A widely used diffusion MRI method for exchange mapping is filter-exchanging imaging (Lasič et al., 2011; Nilsson et al., 2013) which has previously demonstrated ability to differentiate between gliomas and meningiomas (Lampinen et al., 2017). However, FEXI is susceptible to biased exchange estimation due to effects of restricted diffusion and imaging gradients (Lasič et al., 2018; Ohene et al., 2023; Ulloa et al., 2017). A more recent diffusion MRI approach using free gradient waveforms (termed Restriction-Exchange, ResEx) resolves the shortcomings of FEXI and enables more reliable exchange estimation unconfounded by restricted diffusion (Chakwizira et al., 2022; Lasič et al., 2024; Nilsson et al., 2017; Ning et al., 2018). ResEx has been used to measure exchange in the healthy brain but is yet to be applied in pathology (Chakwizira et al., 2023).

In this study, we investigate the potential of exchange mapping using ResEx for characterising gliomas in the human brain. We conduct preoperative assessments of thirteen participants using a clinical MRI scanner with an 80 mT/m gradient strength, with the goal of evaluating the potential value of water exchange rate measurements obtained through diffusion MRI with free gradient waveforms.

## 2 Theory

According to the ResEx framework, the effects of restricted diffusion and exchange manifest on the diffusion-weighted signal, $S$, according to (Chakwizira et al., 2022, 2023)

$$\ln(S/S_0) = -b[E_D + V_\omega E_R] + \frac{1}{2}b^2\left[V_D + 2V_\omega C_{D,R} + V_\omega^2 V_R\right] \cdot (1 - k\Gamma) \tag{1}$$

where we have six microstructural parameters: $S_0$ which is the signal in the absence of diffusion weighting, $E_D$ is the mean free diffusivity of all environments, $E_R$ which is the mean restriction coefficient sensitive to cell size and the fraction of environments undergoing restricted diffusion, $V_D$ and $V_R$ which are the variances in $D$ and $R$, respectively, $C_{D,R}$ which is the covariance between $D$ and $R$ and finally $k$ which is the intercompartmental water exchange rate. The signal equation also features three experimental parameters: $b$ which is the b-value measuring the strength of the diffusion weighting, $V_\omega$ which is the strength of the restriction weighting performed by a given gradient waveform $g(t)$ defined through

$$V_\omega = \frac{\gamma^2}{b}\int_0^T g^2(t)\mathrm{d}t \tag{2}$$

where $\gamma$ is the gyromagnetic ratio, $\Gamma$ is the strength of the exchange-weighting defined through

$$\Gamma = \frac{2}{b^2}\int_0^T tq^2(t)q^2(t+t')\mathrm{d}t' \tag{3}$$

where $q(t) = \gamma \int_0^t g(t')\mathrm{d}t'$ is the dephasing q-vector. The ResEx framework enables a two-dimensional restriction-exchange experiment design, where the acquisition is built to comprise a set of waveforms that independently vary $\Gamma$ and $V_\omega$ to separate restriction effects from exchange.

## 3 Methods

This section outlines the design of the gradient waveforms used to probe water exchange, their implementation on a clinical MRI system, the patient characteristics, and the data preprocessing and analysis steps.

*Gradient waveform design*

To enable independent mapping of exchange and restriction effects, two gradient waveforms were designed following the approach of (Chakwizira et al., 2022) (Fig. 1A-B). The waveforms were differentially sensitive to exchange but similarly sensitive to restricted diffusion, that is, they had different $\Gamma$ (10 and 40 ms) but equal $V_\omega$ (3500 $s^{-2}$). Due to scan time constraints, it was not feasible to also include waveforms with variable sensitivity to restrictions.

Waveforms were one-dimensional (linear tensor encoding), designed to be symmetric about the refocusing pulse, had a maximum encoding time of 120 ms, and a pause duration (time between the end of first and start of second encoding block) of 9 ms, during which the 180 RF pulse was played out. The waveforms all achieved a minimum b-value of 4 ms/$\mu m^2$ at the maximum gradient strength. All waveforms used gradient strength and slew rates below 80 mT/m and 70 T/m/s, respectively.

*Numerical simulations*

To verify that the protocol above was selectively sensitive to exchange (permeability) variations, Monte Carlo simulations were performed in a substrate of spheres in exchange with the extracellular space. The spheres had diameters drawn from a Gamma distribution, with an average diameter of 2 µm. The permeability of the spheres was set to either zero or to yield an exchange rate of 20 $s^{-1}$, following the approach of (Chakwizira et al., 2023). Signals were generated at b-values of 0, 1.3, 2.6 and 4 ms/$\mu m^2$. All simulations were performed using the framework described in (Chakwizira et al., 2022).

*Participants*

The patient group comprised thirteen participants with primary brain tumours, aged between 43 and 77 years. Ten participants were diagnosed with grade 4 glioblastoma, two with grade 3 astrocytoma and one with grade 4 astrocytoma. The study was conducted in accordance with the Declaration of Helsinki, and approved by the Regional Ethics Board Lund, Sweden and by the Swedish Ethical Review Authority prior to starting any work. Informed consent was obtained from all subjects involved in the study. The inclusion criteria were as follows: above 18 years old, undergone biopsy or resection of the brain lesion for pathological analysis, and completion of preoperative MRI examination, including conventional MRI sequences. Exclusion criteria were having a pacemaker or metallic wires in the body incompatible with MRI and being unable to sign informed consent.

*MRI scans*

All thirteen participants were examined on a 3 T MAGNETOM Prisma (Siemens Healthineers, Forchheim, Germany) with a research sequence (Szczepankiewicz et al., 2019) using the gradient waveforms described above. For each waveform, images were acquired at b-values of 0, 1.3, 2.6, and 4.0 ms/μm$^2$, with 10 directions at each b-value, chosen to minimise electrostatic repulsion on a sphere. Spatial resolution was set to 2×2 mm$^2$ in-plane and a slice thickness of 5 mm. Other imaging parameters were TE = 138 ms, TR = 4.6 s, strong fat suppression, and a total acquisition time of 2.5 minutes per waveform, giving a total scan time of 5 minutes. In addition to the ResEx scans, pre and post gadolinium T1-weighted and FLAIR images were also acquired at 1 mm isotropic resolution in all participants. One of the participants was examined at five different time-points, allowing for a longitudinal analysis.

*Data analysis*

Diffusion-weighted images were denoised using the Machenko-Pastur principal component analysis (Veraart et al., 2016) (github.com/Neurophysics-CFIN/MP-PCA-Denoising), after which they were corrected for motion and eddy currents using Elastix (Klein et al., 2010) with extrapolated references (Nilsson et al, 2016). The images were then powder-averaged for further analysis. To allow comparison between exchange measurements and contrast enhancement, diffusion-weighted images for each participant were rigidly coregistered with the corresponding T1-weighted volume. FLAIR images were also coregistered to the T1-weighted volume to aid delineation of oedema.

An initial analysis of exchange effects in the gliomas was performed by studying the exchange-driven contrast between the two waveforms at different b-values. The contrast was calculated through the normalised exchange-driven contrast

$$\Delta S(b) = \frac{\left(S_{low\,\Gamma}(b) - S_{high\,\Gamma}(b)\right)}{\frac{1}{2}\left(S_{low\,\Gamma}(b) + S_{high\,\Gamma}(b)\right)} \tag{4}$$

where $S_{low\,\Gamma}(b)$ and $S_{high\,\Gamma}(b)$ are the signals acquired using the waveforms with the lower and higher values of $\Gamma$, respectively, at the same b-value.

ResEx parameter estimates were obtained via the voxelwise fitting of Eq. (1) using the non-linear least squares solver *lsqnonlin* in MATLAB (The MathWorks, Natick, MA, R2024a). Since the scans did not include restriction-encoding waveforms, fitting was performed with the restriction-related parameters in Eq. 1 set to zero, yielding

$$\ln(S/S_0) = -b \cdot E_D + \frac{1}{6} b^2 E_D^2 K_T \cdot (1 - k\Gamma) \tag{5}$$

where $K_T = 3V_D/E_D^2$ is the kurtosis. The fitting produced maps of the apparent diffusion coefficient ($E_D$), kurtosis and the exchange rate.

Explorative ROI-based analyses were also performed after tumour segmentation using the BraTS algorithm (Ferreira et al., 2024) and the BRATS orchestator (Kofler et al., 2025) with the glioma pre- and post-treatment segmentor (Jain et al., 2025), which delineated three mutually exclusive regions of interest: oedema, tumour core, and contrast-enhancing parts of the tumour. Contralateral healthy grey and white matter were also segmented using SynthSeg (Billot et al., 2023). Quantitative comparisons of exchange estimates in the tumour versus healthy white and grey matter, as well as tumour versus oedema were performed. In each case, significance was verified using a two-tailed paired t-test with a significance threshold of 0.05.

## 4 Results

Figure 1 shows an overview of the expected effects of the two exchange-encoding waveforms in different types of microstructures. In the presence of restricted diffusion without exchange, the waveforms produce highly similar signal-versus-b curves with negligible contrast. Introducing exchange gives rise to notable contrast between the signals from the two waveforms, with the effect becoming more pronounced at higher b-values. This illustrates that the two waveforms are specific to exchange and that exchange is, as expected, a higher-order effect (appears at higher b-values). These same patterns were also observed in a grade IV glioblastoma (panel C). The lesion identified on post-Gd

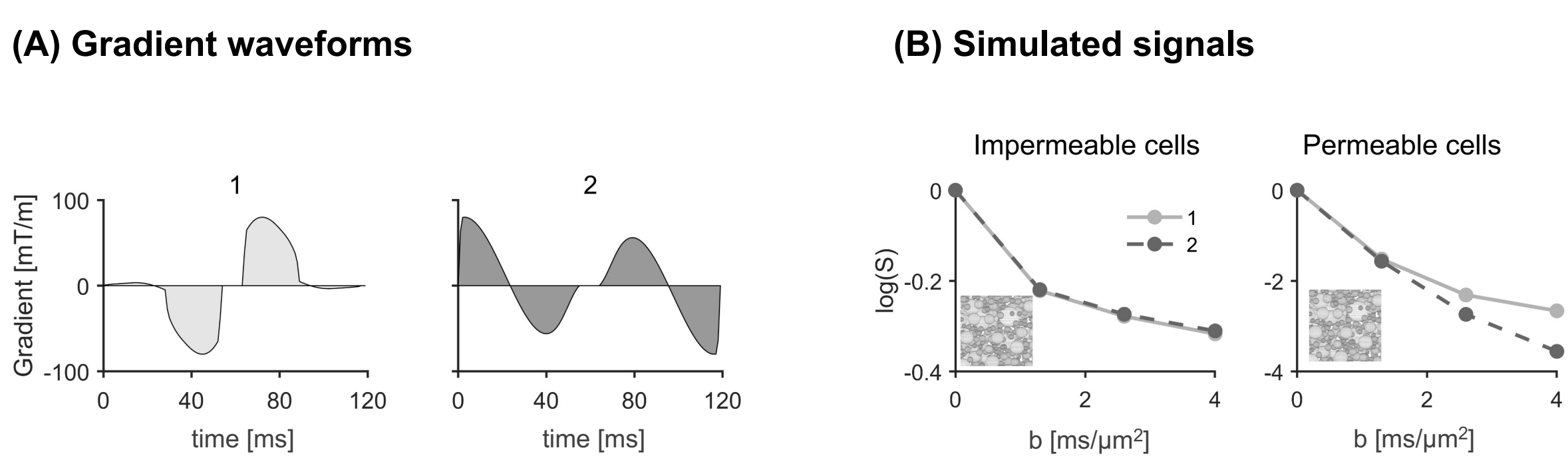

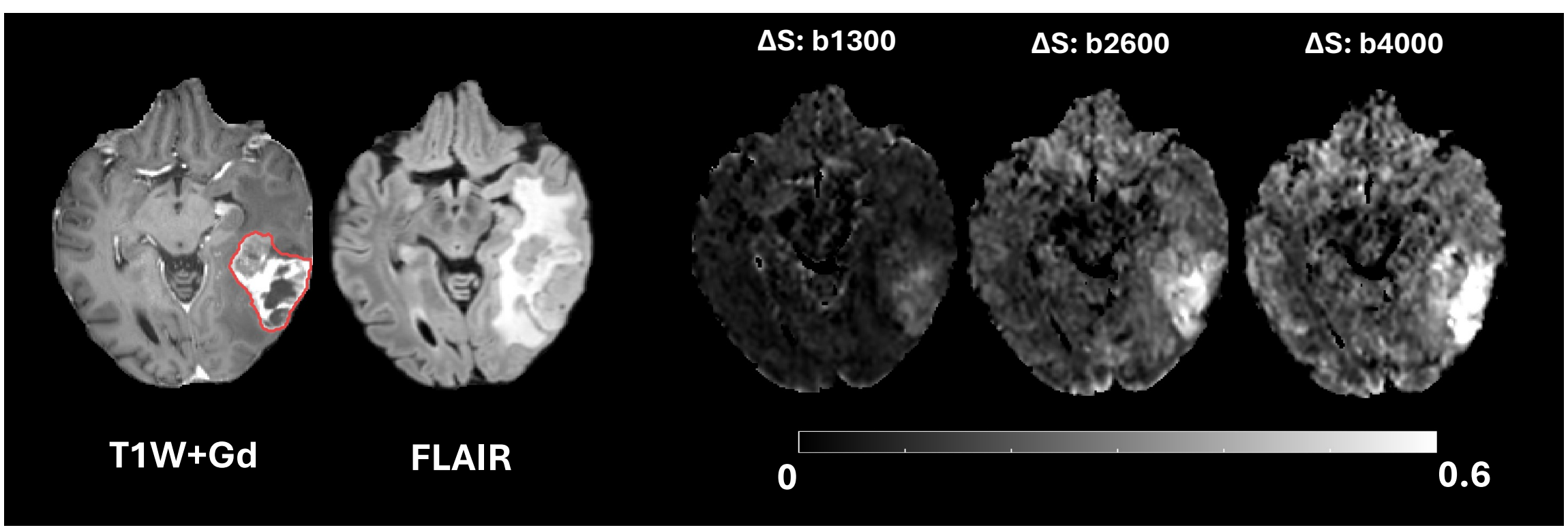

Figure 1: Exchange manifesting as signal differences. (A): Two different free gradient waveforms were used in the study, which differed in their exchange weighting (Γ), but featured similar degrees of restriction weighting ($V_\omega$ ). (B): Simulated signal-versus-b curves for each waveform, showing negligible contrast in impermeable cells but strong signal difference at high b-values for permeable cells. This illustrates the specificity of the approach to exchange. (C). In vivo demonstration of the exchange-driven signal contrast (ΔS is the difference in signals obtained with the two waveforms) at different b-values in a grade IV glioma, shown alongside post-Gd T1-weighted and FLAIR images.

T1 and FLAIR images shows corresponding elevated exchange-driven contrast ($\Delta S$). The contrast is more conspicuous at higher b-values, indicating that it is driven by a higher-order time-dependent effect. Interestingly, as the b-value increase, the contrast varies not only in magnitude but also spatially, more clearly delineating the tumour region. Higher $\Delta S$ values are concentrated in the enhancing lesion and surrounding abnormal tissue. Normal-appearing brain parenchyma exhibits low $\Delta S$ across all b-values. Notably, the large region of FLAIR hyperintensity surrounding the tumour (likely vasogenic oedema) is not hyperintense on the exchange-driven contrast. This suggests that the exchange contrast responds to true microstructural changes, rather than merely to water accumulation.

Figure 2 presents ResEx parameter estimates derived from fitting Equation 5 to the acquired data, alongside post-Gd T1-weighted and FLAIR images for anatomical reference. Results are shown for five representative cases (three glioblastomas and two astrocytomas), in the slices where the tumour had maximum in-plane extent. The T1W images highlight the enhancing tumour (red ROIs) while FLAIR shows extensive peritumoural hyperintensity consistent with oedema. Regions of FLAIR hyperintensity generally coincide with elevated ADC and reduced kurtosis, which suggests increased water mobility, low cellularity, and reduced microstructural heterogeneity.

The exchange maps in Fig. 2 show elevated water exchange rates within tumour tissue across all cases considered. However, in some cases (third glioblastoma and first astrocytoma), exchange hyperintensity appears to extend well beyond the tumour delineated by the post-Gd T1, suggesting that the exchange maps provide information captured by neither the morphological images nor conventional diffusion MRI maps. Regions of highly elevated ADC (fluid-filled or necrotic) are associated with very low exchange rates. Exchange estimates in these areas are, however, less reliable since the kurtosis is very low. When kurtosis approaches zero, exchange estimation becomes unstable because exchange is inferred from the temporal evolution of kurtosis (Eq. 5).

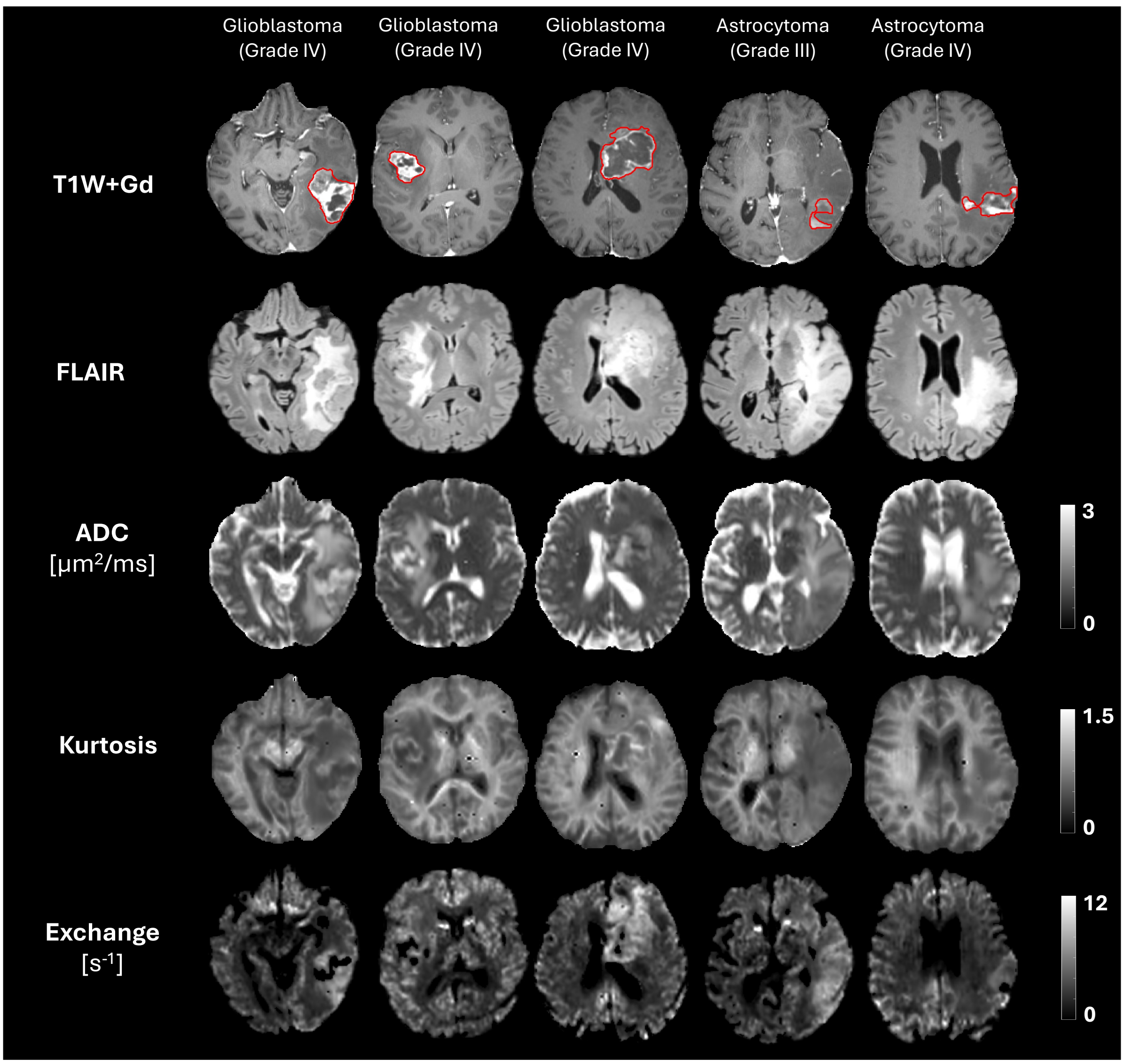

Figure 2: Water exchange estimation in gliomas. Post-Gadolinium T1-weighted images are shown alongside FLAIR in three glioblastoma and two astrocytoma patients. Red ROIs demarcate whole tumour on the T1-weighted images. FLAIR hyperintensity generally colocalises with elevated ADC and reduced kurtosis. Tumours consistently show elevated exchange rates, save for high-ADC, likely fluid-filled regions. Exchange elevation is evident beyond the enhancing tumour in some cases, likely reflecting oedema, although this pattern is not reflected across all shown cases. The exchange maps provide information complementary to that available in the other maps. Note that a signal loss appearing in a U-shaped form in the posterior part of the brain is due to a fat artifact in the diffusion-weighted images.

Figure 3 presents a quantitative comparison of the exchange rates in the tumour to those in contralateral healthy grey and white matter, across all thirteen participants included in the study. Paired comparisons reflect a systematic increase in exchange rate in tumour tissue relative to both healthy white and grey matter in almost all participants, reinforcing the elevation of exchange in the tumour microenvironment already suggested by Fig. 2. Group-level analyses also confirmed that exchange rates were significantly higher in the tumour compared to healthy white matter (p = 0.0002) and grey matter (p = 0.0005). Exchange elevation likely signifies increased cell membrane turnover or metabolic activity, suggesting the potential of exchange estimation to distinguish tumour tissue from healthy brain.

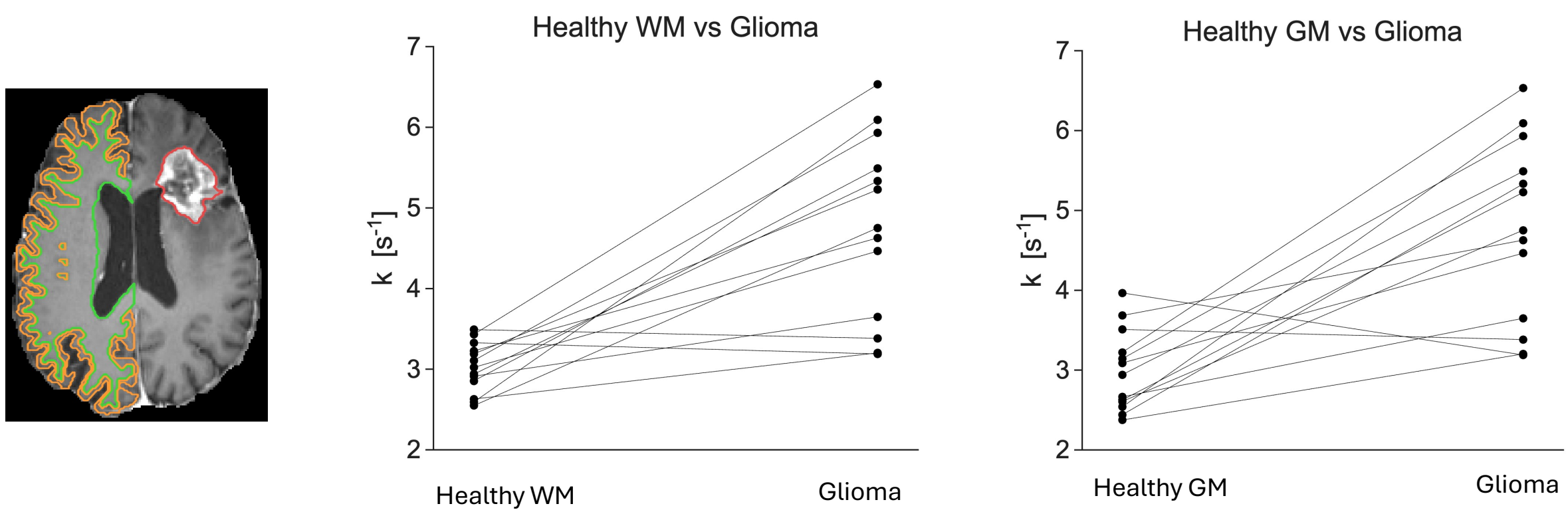

Figure 3: Comparison of exchange rates between gliomas and contralateral healthy brain tissue in thirteen participants. Leftmost panel shows an example post-Gd T1-weighted image reflecting the tumour (red), healthy white matter (green) and healthy grey matter (yellow). Exchange rates are significantly higher in gliomas compared to both healthy white matter (p = 0.0002) and grey matter (p = 0.0005).

The exchange maps displayed in Fig. 2 interestingly do not clearly differentiate between oedema and normal-appearing tissue. However, they generally appear highly sensitive to the boundary between oedema and tumour, which suggests that exchange rate mapping may serve as a useful predictor of resection margins in surgical planning. A quantitative comparison of the exchange rates in peritumoural oedema and those in healthy white matter and enhancing tumour is presented in Fig. 4, across all thirteen participants. Figure 4 reveals that exchange rates in oedema are generally higher than those observed in healthy white matter in most participants (p = 0.02) but lower than those observed in the enhancing tumour in all participants (p = 0.00004). The elevation of exchange rates in oedema relative to healthy tissue suggests that, in some individuals, the oedematous region may be associated with significant disruption of membrane integrity or increased permeability, potentially reflecting underlying processes such as inflammation, necrosis, or active tumour infiltration. The clear distinction between exchange estimates in oedema and enhancing tumour supports the previous proposition that the water exchange rate may facilitate the distinction between tumour tissue and oedematous regions.

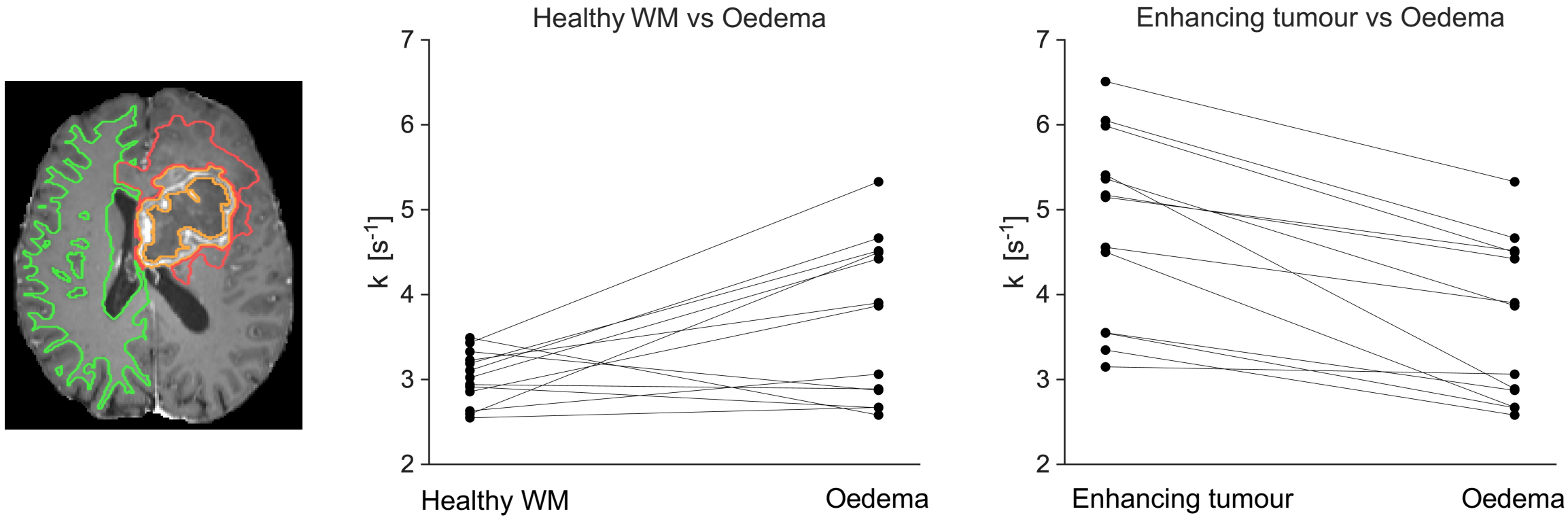

Figure 4: Comparison of exchange rates between gliomas and oedema in thirteen participants. Leftmost panel shows a post-Gd T1-weighted image from one participant reflecting oedema (red), enhancing tumour (yellow) and healthy white matter (green). Exchange rates are significantly higher in oedema compared to healthy white matter ($p = 0.02$) and significantly lower in oedema compared to the enhancing tumour ($p = 0.00004$).

Figure 5 shows ResEx parameter maps alongside post-contrast T1-weighted and FLAIR images from the longitudinal assessment of one participant diagnosed with grade 4 glioblastoma. The T1-weighted images reveal progressive enlargement of both the core and the enhancing tumour region over time, while the FLAIR images show persistent and evolving peritumoural hyperintensity. The tumour ROI defined on the T1 images expands across successive scans, indicating disease progression. ADC maps consistently show hyperintensity in the tumour and surrounding abnormal tissue, indicative of oedema, fluid likely from previous resection, and sustained necrotic components. By contrast, kurtosis maps show elevated values predominantly in the enhancing rim, consistent with increasing microstructural complexity and tumour progression. Exchange maps further support this interpretation: while the core remains low in exchange, the surrounding rim shows elevated and progressively increasing exchange rates, matching the gadolinium enhancement pattern. This rise in exchange may reflect worsening membrane integrity or upregulation of AQP4 expression, both associated with aggressive tumour behaviour. Notably, exchange hyperintensity appears to extend beyond the enhancing margin, in particular in the posterior direction, potentially suggesting a growing degree of infiltrative tumour spread into peritumoural tissue.

Overall, Fig. 5 indicates that the exchange rate measured with diffusion MRI remains spatially associated with the direction of tumour growth throughout disease progression.

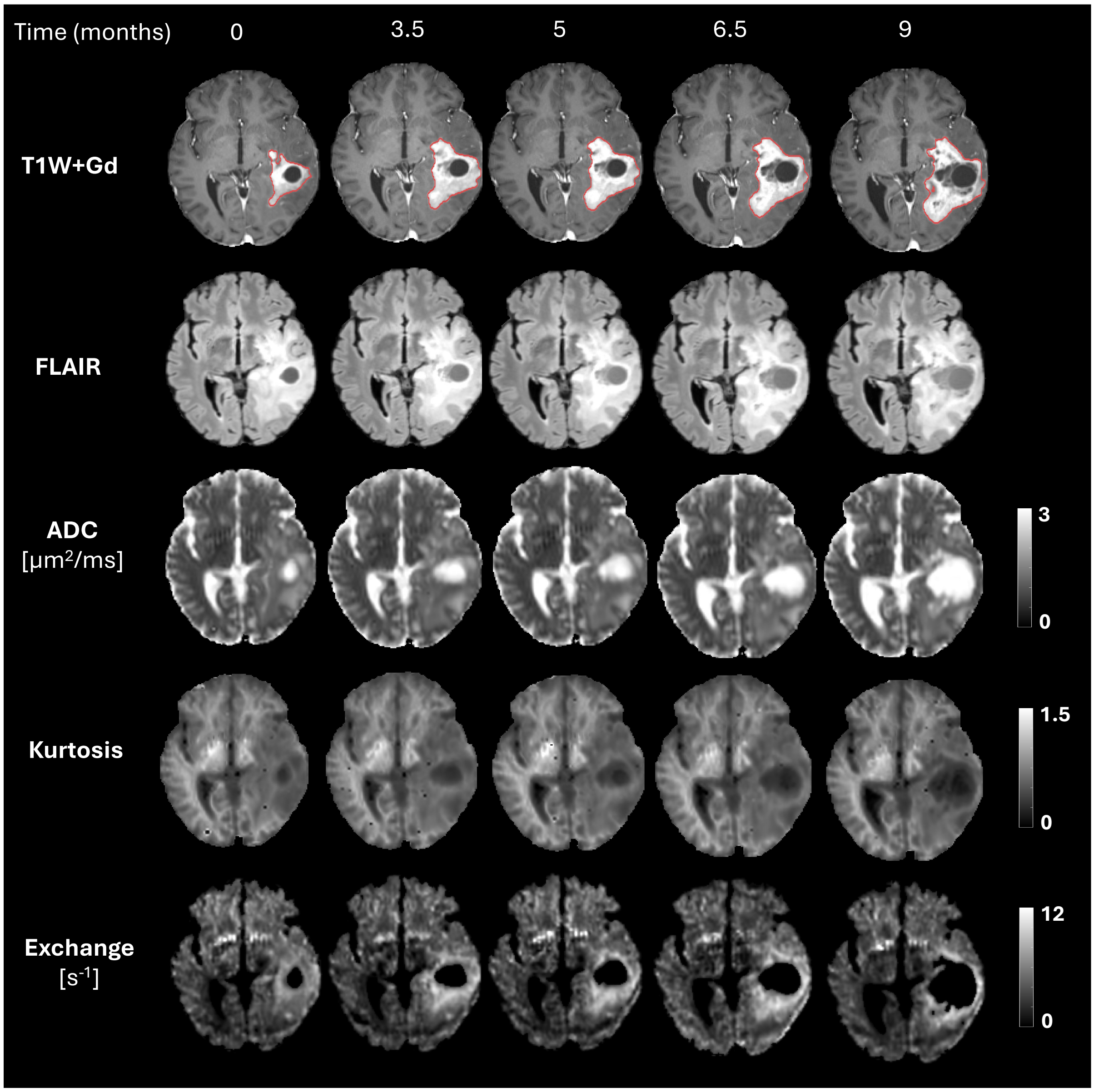

Figure 5: Longitudinal ResEx parameter maps from a glioblastoma grade IV patient, showing tumour progression over a nine-month period. Post-Gd T1-weighted, FLAIR, ADC, kurtosis, and exchange maps reveal temporal changes, with exchange hyperintensity remaining largely consistent with the pattern of Gd enhancement.

## 5 Discussion

This study demonstrates that elevated water exchange, as measured by diffusion MRI with free gradient waveforms, can be detected in gliomas. The exchange rates in most tumours were considerably higher than those in healthy white and grey matter, in which the rates were in the same range as those observed previously using this method (Chakwizira et al., 2023) (Fig. 2-3). The variability in exchange effects was not consistently predicted by gadolinium enhancement, suggesting that exchange imaging provides complementary information to conventional contrast-enhanced MRI (Fig. 2). Oedema generally showed low exchange rates relative to enhancing tumour, but slightly elevated rates relative to healthy white matter (Fig. 4). Furthermore, from a visual inspection, some patients displayed heterogeneity in exchange rates within the oedema. This suggests that exchange contrast is not simply a reflection of vasogenic water content, as visualized with FLAIR, but instead reflects more subtle membrane-related physiology. Necrotic regions were characterised by high ADC, low kurtosis, and negligible exchange, consistent with loss of membrane integrity and structural disintegration. Collectively, our results highlight the spatial and individual heterogeneity of gliomas and suggest that diffusion-based exchange mapping reveals physiological features beyond those accessible with standard anatomical imaging or conventional diffusion MRI metrics.

The presence of elevated exchange in tumour regions may reflect increased membrane permeability, possibly mediated by upregulation of aquaporin-4 (AQP4) or other structural and metabolic alterations. Studies have implicated AQP4 in glioma progression and shown a link between AQP4 expression and water diffusivity and kurtosis in human astrocytomas (Tan et al., 2016). Elevated AQP4 and increased surface-to-volume ratio due to cellular proliferation or microvascular remodelling may contribute to the increased water exchange we observed. Disruption of the blood-brain barrier (BBB) may also allow for enhanced intercompartmental water exchange (Wang et al., 2023). However, it is unlikely that BBB exchange is strongly linked to our results due to the high b-values that we used, which should suppress the vascular component of the signal. Interestingly, exchange was elevated around the tumour periphery, even beyond the enhancing margin, potentially suggesting sensitivity to the main direction of infiltrative tumour growth (Fig. 2, Fig. 5). While speculative, this points to a potential role for exchange-sensitive diffusion MRI in planning surgical margins or radiotherapy. At a minimum, our results clearly establish water exchange as a distinct contrast mechanism that warrants further investigation in glioma imaging.

These findings have both methodological and clinical implications. Water exchange alters the diffusion-weighted signal in ways not accounted for by standard diffusion models, such as those used in diffusion kurtosis imaging (DKI). This contribution is likely to confound the interpretation of data. This is particularly relevant because DKI has been shown to assist in glioma grading, with several studies reporting significantly higher kurtosis values in high-grade compared to low-grade gliomas (Abdalla et al., 2020; Delgado et al., 2017). However, since exchange can influence kurtosis estimates, accounting for exchange effects may improve the reliability and biological specificity of DKI. Incorporating waveforms that vary in exchange sensitivity into clinical protocols may thus improve the precision of diffusion MRI and refine biophysical modeling of tumour microstructure.

Nonetheless, limitations remain. An important one is the potential for poor generalization due to a relatively small sample. Furthermore, exchange rates estimated with diffusion MRI are apparent values that depend on the experimental design and model assumptions. While we minimized the influence of restricted diffusion by design, the presence of such effects may still bias exchange estimates if not properly controlled. Moreover, the approach is best suited to high gradient strengths, limiting its use to high-end MRI scanners. Despite these constraints, our findings support the utility of exchange mapping with ResEx as a tool for non-invasive glioma characterisation.

In conclusion, this work provides motivation for further exploration of water exchange imaging in clinical settings for tumour diagnosis, grading, and treatment monitoring.

## Acknowledgements

We thank Siemens Healthineers for access to the pulse sequence programming environment.

## Data and code availability statement

The ROI data and code supporting the findings of this study will be made openly available in the repository at https://github.com/arthur-chakwizira/resex-glioma

## Declaration of Competing Interests

Markus Nilsson, Carl-Fredrik Westin, and Filip Szczepankiewicz have financial conflicts of interest related to patents in related methods. Arthur Chakwizira, Linda Knutsson, and Pia C Sundgren have no conflicts of interest to disclose.

## CRediT authorship contribution statement

Arthur Chakwizira: Conceptualisation, Formal Analysis, Investigation, Methodology, Software, Validation, Visualisation, Writing – Original Draft Preparation

Filip Szczepankiewicz: Supervision, Funding Acquisition, Software, Writing – Review & Editing

Carl-Fredrik Westin: Supervision, Funding Acquisition, Writing – Review & Editing

Linda Knutsson: Supervision, Writing – Review & Editing

Pia C Sundgren: Supervision, Funding Acquisition, Writing – Review & Editing

Markus Nilsson: Conceptualisation, Funding Acquisition, Software, Methodology, Supervision, Writing – Review & Editing

## References

Abdalla, G., Dixon, L., Sanverdi, E., Machado, P. M., Kwong, J. S. W., Panovska-Griffiths, J., Rojas-Garcia, A., Yoneoka, D., Veraart, J., Van Cauter, S., Abdel-Khalek, A. M., Settein, M., Yousry, T., & Bisdas, S. (2020). The diagnostic role of diffusional kurtosis imaging in glioma grading and differentiation of gliomas from other intra-axial brain tumours: A systematic review with critical appraisal and meta-analysis. *Neuroradiology*, *62*(7), 791–802. https://doi.org/10.1007/s00234-020-02425-9

Andrasko, J. (1976). Water diffusion permeability of human erythrocytes studied by a pulsed gradient NMR technique. *Biochimica et Biophysica Acta (BBA) - General Subjects*, *428*(2), 304–311. https://doi.org/10.1016/0304-4165(76)90038-6

Billot, B., Greve, D. N., Puonti, O., Thielscher, A., Van Leemput, K., Fischl, B., Dalca, A. V., & Iglesias, J. E. (2023). SynthSeg: Segmentation of brain MRI scans of any contrast and resolution without retraining. *Medical Image Analysis*, *86*, 102789. https://doi.org/10.1016/j.media.2023.102789

Blocher, J., Eckert, I., Elster, J., Wiefek, J., Eiffert, H., & Schmidt, H. (2011). Aquaporins AQP1 and AQP4 in the cerebrospinal fluid of bacterial meningitis patients. *Neuroscience Letters*, *504*(1), 23–27. https://doi.org/10.1016/j.neulet.2011.08.049

Castañeyra-Ruiz, L., González-Marrero, I., González-Toledo, J. M., Castañeyra-Ruiz, A., de Paz-Carmona, H., Castañeyra-Perdomo, A., & Carmona-Calero, E. M. (2013). Aquaporin-4 expression in the cerebrospinal fluid in congenital human hydrocephalus. *Fluids and Barriers of the CNS*, *10*(1), 18. https://doi.org/10.1186/2045-8118-10-18

Chakwizira, A., Westin, C.-F., Brabec, J., Lasič, S., Knutsson, L., Szczepankiewicz, F., & Nilsson, M. (2022). Diffusion MRI with pulsed and free gradient waveforms: Effects of restricted diffusion and exchange. *NMR in Biomedicine*, *n/a*(n/a), e4827. https://doi.org/10.1002/nbm.4827

Chakwizira, A., Zhu, A., Foo, T., Westin, C.-F., Szczepankiewicz, F., & Nilsson, M. (2023). Diffusion MRI with free gradient waveforms on a high-performance gradient system: Probing restriction and exchange in the human brain. *NeuroImage*, *283*, 120409. https://doi.org/10.1016/j.neuroimage.2023.120409

Delgado, A. F., Fahlström, M., Nilsson, M., Berntsson, S. G., Zetterling, M., Libard, S., Alafuzoff, I., van Westen, D., Lätt, J., Smits, A., & Larsson, E.-M. (2017). Diffusion Kurtosis Imaging of Gliomas Grades II and III - A Study of Perilesional Tumor Infiltration, Tumor Grades and Subtypes at Clinical Presentation. *Radiology and Oncology*, *51*(2), 121–129. https://doi.org/10.1515/raon-2017-0010

Ferreira, A., Solak, N., Li, J., Dammann, P., Kleesiek, J., Alves, V., & Egger, J. (2024). *How we won BraTS 2023 Adult Glioma challenge? Just faking it! Enhanced Synthetic Data Augmentation and Model Ensemble for brain tumour segmentation* (arXiv:2402.17317). arXiv. https://doi.org/10.48550/arXiv.2402.17317

Jain, I., Willems, S., Latre, S., & Schepper, T. D. (2025). *On-the-Fly Data Augmentation for Brain Tumor Segmentation* (arXiv:2509.24973). arXiv. https://doi.org/10.48550/arXiv.2509.24973

Jia, Y., Xu, S., Han, G., Wang, B., Wang, Z., Lan, C., Zhao, P., Gao, M., Zhang, Y., Jiang, W., Qiu, B., Liu, R., Hsu, Y.-C., Sun, Y., Liu, C., Liu, Y., & Bai, R. (2023). Transmembrane water-efflux rate measured by magnetic resonance imaging as a biomarker of the expression of aquaporin-4 in gliomas. *Nature Biomedical Engineering*, *7*(3), 236–252. https://doi.org/10.1038/s41551-022-00960-9

Kärger, J. (1985). NMR self-diffusion studies in heterogeneous systems. *Advances in Colloid and Interface Science*, *23*, 129–148. https://doi.org/10.1016/0001-8686(85)80018-X

Klein, S., Staring, M., Murphy, K., Viergever, M. A., & Pluim, J. P. W. (2010). elastix: A Toolbox for Intensity-Based Medical Image Registration. *IEEE Transactions on Medical Imaging*, *29*(1), 196–205. IEEE Transactions on Medical Imaging. https://doi.org/10.1109/TMI.2009.2035616

Kofler, F., Rosier, M., Astaraki, M., Baid, U., Möller, H., Buchner, J. A., Steinbauer, F., Oswald, E., Rosa, E. de la, Ezhov, I., See, C. von, Kirschke, J., Schmick, A., Pati, S., Linardos, A., Pitarch, C., Adap, S., Rudie, J., Verdier, M. C. de, … Menze, B. (2025). *BraTS orchestrator: Democratizing and Disseminating state-of-the-art brain tumor image analysis* (arXiv:2506.13807). arXiv. https://doi.org/10.48550/arXiv.2506.13807

Lampinen, B., Szczepankiewicz, F., van Westen, D., Englund, E., C Sundgren, P., Lätt, J., Ståhlberg, F., & Nilsson, M. (2017). Optimal experimental design for filter exchange imaging: Apparent exchange rate measurements in the healthy brain and in intracranial tumors. *Magnetic Resonance in Medicine*, *77*(3), 1104–1114. https://doi.org/10.1002/mrm.26195

Lasič, S., Chakwizira, A., Lundell, H., Westin, C.-F., & Nilsson, M. (2024). Tuned exchange imaging: Can the filter exchange imaging pulse sequence be adapted for applications with thin slices and restricted diffusion? *NMR in Biomedicine*, *37*(11), e5208. https://doi.org/10.1002/nbm.5208

Lasič, S., Lundell, H., Topgaard, D., & Dyrby, T. B. (2018). Effects of imaging gradients in sequences with varying longitudinal storage time-Case of diffusion exchange imaging. *Magnetic Resonance in Medicine*, *79*(4), 2228–2235. https://doi.org/10.1002/mrm.26856

Lasič, S., Nilsson, M., Lätt, J., Ståhlberg, F., & Topgaard, D. (2011). Apparent exchange rate mapping with diffusion MRI. *Magnetic Resonance in Medicine*, *66*(2), 356–365. https://doi.org/10.1002/mrm.22782

Montgomery, M. K., Kim, S. H., Dovas, A., Zhao, H. T., Goldberg, A. R., Xu, W., Yagielski, A. J., Cambareri, M. K., Patel, K. B., Mela, A., Humala, N., Thibodeaux, D. N., Shaik, M. A., Ma, Y., Grinband, J., Chow, D. S., Schevon, C., Canoll, P., & Hillman, E. M. C. (2020). Glioma-Induced Alterations in Neuronal Activity and Neurovascular Coupling during Disease Progression. *Cell Reports*, *31*(2), 107500. https://doi.org/10.1016/j.celrep.2020.03.064

Nayak, L., & Reardon, D. A. (2017). High-grade Gliomas. *Continuum (Minneapolis, Minn.)*, *23*(6, Neuro-oncology), 1548–1563. https://doi.org/10.1212/CON.0000000000000554

Nico, B., Mangieri, D., Tamma, R., Longo, V., Annese, T., Crivellato, E., Pollo, B., Maderna, E., Ribatti, D., & Salmaggi, A. (2009). Aquaporin-4 contributes to the resolution of peritumoural brain oedema in human glioblastoma multiforme after combined chemotherapy and radiotherapy. *European Journal of Cancer (Oxford, England: 1990)*, *45*(18), 3315–3325. https://doi.org/10.1016/j.ejca.2009.09.023

Nilsson, M., Lasič, S., Drobnjak, I., Topgaard, D., & Westin, C.-F. (2017). Resolution limit of cylinder diameter estimation by diffusion MRI: The impact of gradient waveform and orientation dispersion. *NMR in Biomedicine*, *30*(7), e3711. https://doi.org/10.1002/nbm.3711

Nilsson, M., Lätt, J., Westen, D. van, Brockstedt, S., Lasič, S., Ståhlberg, F., & Topgaard, D. (2013). Noninvasive mapping of water diffusional exchange in the human brain using filter-exchange imaging. *Magnetic Resonance in Medicine*, *69*(6), 1572–1580. https://doi.org/10.1002/mrm.24395

Ning, L., Nilsson, M., Lasič, S., Westin, C.-F., & Rathi, Y. (2018). Cumulant expansions for measuring water exchange using diffusion MRI. *The Journal of Chemical Physics*, *148*(7), 074109. https://doi.org/10.1063/1.5014044

Ohene, Y., Harris, W. J., Powell, E., Wycech, N. W., Smethers, K. F., Lasič, S., South, K., Coutts, G., Sharp, A., Lawrence, C. B., Boutin, H., Parker, G. J. M., Parkes, L. M., & Dickie, B. R. (2023). Filter exchange imaging with crusher gradient modelling detects increased blood–brain barrier water permeability in response to mild lung infection. *Fluids and Barriers of the CNS*, *20*, 25. https://doi.org/10.1186/s12987-023-00422-7

Omuro, A., & DeAngelis, L. M. (2013). Glioblastoma and other malignant gliomas: A clinical review. *JAMA*, *310*(17), 1842–1850. https://doi.org/10.1001/jama.2013.280319

Ostrom, Q. T., Bauchet, L., Davis, F. G., Deltour, I., Fisher, J. L., Langer, C. E., Pekmezci, M., Schwartzbaum, J. A., Turner, M. C., Walsh, K. M., Wrensch, M. R., & Barnholtz-Sloan, J. S. (2014). The epidemiology of glioma in adults: A "state of the science" review. *Neuro-Oncology*, *16*(7), 896–913. https://doi.org/10.1093/neuonc/nou087

Papadopoulos, M. C., & Verkman, A. S. (2013). Aquaporin water channels in the nervous system. *Nature Reviews. Neuroscience*, *14*(4), 265–277. https://doi.org/10.1038/nrn3468

Sun, D.-P., Lee, Y.-W., Chen, J.-T., Lin, Y.-W., & Chen, R.-M. (2020). The Bradykinin-BDKRB1 Axis Regulates Aquaporin 4 Gene Expression and Consequential Migration and Invasion of Malignant Glioblastoma Cells via a Ca2+-MEK1-

ERK1/2-NF-κB Mechanism. *Cancers*, *12*(3), 667. https://doi.org/10.3390/cancers12030667

Szczepankiewicz, F., Sjölund, J., Ståhlberg, F., Lätt, J., & Nilsson, M. (2019). Tensor-valued diffusion encoding for diffusional variance decomposition (DIVIDE): Technical feasibility in clinical MRI systems. *PLOS ONE*, *14*(3), e0214238. https://doi.org/10.1371/journal.pone.0214238

Tan, Y., Zhang, H., Zhao, R.-F., Wang, X.-C., Qin, J.-B., & Wu, X.-F. (2016). Comparison of the values of MRI diffusion kurtosis imaging and diffusion tensor imaging in cerebral astrocytoma grading and their association with aquaporin-4. *Neurology India*, *64*(2), 265–272. https://doi.org/10.4103/0028-3886.177621

Ulloa, P., Methot, V., & Koch, M. A. (2017). Experimental validation of a bias in apparent exchange rate measurement. *Current Directions in Biomedical Engineering*, *3*(2), 529–532. https://doi.org/10.1515/cdbme-2017-0112

Veraart, J., Novikov, D. S., Christiaens, D., Ades-aron, B., Sijbers, J., & Fieremans, E. (2016). Denoising of diffusion MRI using random matrix theory. *NeuroImage*, *142*, 394. https://doi.org/10.1016/j.neuroimage.2016.08.016

Verkman, A. S., Hara-Chikuma, M., & Papadopoulos, M. C. (2008). Aquaporins—New players in cancer biology. *Journal of Molecular Medicine (Berlin, Germany)*, *86*(5), 523–529. https://doi.org/10.1007/s00109-008-0303-9

Wang, Z., Wang, B., Li, Z., Han, G., Meng, C., Jiao, B., Guo, K., Hsu, Y.-C., Sun, Y., Liu, Y., & Bai, R. (2023). The Consistence of Dynamic Contrast-Enhanced MRI and Filter-Exchange Imaging in Measuring Water Exchange Across the Blood-Brain Barrier in High-Grade Glioma. *Journal of Magnetic Resonance Imaging: JMRI*, *58*(6), 1850–1860. https://doi.org/10.1002/jmri.28729

Yang, B., & Verkman, A. S. (1997). Water and glycerol permeabilities of aquaporins 1-5 and MIP determined quantitatively by expression of epitope-tagged constructs in Xenopus oocytes. *The Journal of Biological Chemistry*, *272*(26), 16140–16146. https://doi.org/10.1074/jbc.272.26.16140